\documentclass[a4paper,12pt]{article}

\usepackage[T1]{fontenc}
\usepackage{psfig}
\usepackage{amssymb}
\usepackage{amsmath}

\def\ZZ{{\mathchoice {\mathsf{Z \hspace{-0.45em} Z}} {\mathsf{Z 
        \hspace{-0.45em} Z}} {\mathsf{Z \hspace{-0.32em} Z}} 
    {\mathsf{Z \hspace{-0.23em} Z}}}}

\def\RR{{\mathchoice {\mathrm{I \hspace{-0.2em} R}} {\mathrm{I 
        \hspace{-0.2em} R}} {\mathrm{I \hspace{-0.14em} R}}
    {\mathrm{I \hspace{-0.14em} R}}}}

\def\PP{{\mathchoice {\mathrm{I \hspace{-0.2em} P}} {\mathrm{I 
        \hspace{-0.2em} P}} {\mathrm{I \hspace{-0.14em} P}}
    {\mathrm{I \hspace{-0.14em} P}}}}
\def\EE{{\mathchoice {\mathrm{I \hspace{-0.2em} E}} {\mathrm{I 
        \hspace{-0.2em} E}} {\mathrm{I \hspace{-0.14em} E}}
    {\mathrm{I \hspace{-0.14em} E}}}} 
\def\DD{{\mathchoice {\mathrm{I \hspace{-0.2em} D}} {\mathrm{I 
        \hspace{-0.2em} D}} {\mathrm{I \hspace{-0.14em} D}}
    {\mathrm{I \hspace{-0.14em} D}}}}

\def\tr{\mathrm{tr}}

\def\Ccal{\mathcal{C}}

\newcommand{\lbar}[1]{\underline{#1}}

\newtheorem{theorem}{Theorem}
\newtheorem{lemma}{Lemma}
\newenvironment{proof}[1][{}]{\noindent 
\textbf{Proof #1} \par}{\hfill $\blacksquare$ \par}
\newcounter{rqlfig}

\title{Ward type identities for the 2d Anderson model \\ 
at weak disorder}
\author{Jacques Magnen, Gilles Poirot, Vincent Rivasseau\\
\protect \normalsize 
Centre de Physique Th{\'e}orique, {\'E}cole Polytechnique\\
\protect \normalsize 
91128 Palaiseau Cedex, FRANCE}
\date{}

\begin{document}

\maketitle

\thispagestyle{headings}
\catcode`\@=11
\def\ps@headings{\def\@oddhead{\hfill \begin{tabular}{r} 
\textsc{{\'E}cole Polytechnique} \\ CPHT S 599.0198
\end{tabular}}}
\catcode`\@=12
\relax

\begin{abstract}
Using the particular momentum conservation laws in dimension $d=2$, we can
rewrite the Anderson model in terms of low momentum long range fields, at the
price of introducing electron loops. The corresponding loops
satisfy a Ward type identity, hence are much smaller than expected. This fact
should be useful for a study of the weak-coupling model in the middle of the
spectrum of the free Hamiltonian. 
\end{abstract}


\section{Introduction}
We consider a continuous Anderson model in dimension $d=2$. The model is
defined through the following Hamiltonian 
\begin{equation}
H = -\Delta + \lambda V
\end{equation}
where $V$ is a Gaussian random field which is a regularized white noise. 

We are interested in the density of states at weak disorder ($\lambda \ll 1$)
and in the free spectrum (at energy $E>0$). It is well known \cite{PF} 
that because of ergodicity, the density of states is a
deterministic quantity given by  
\begin{equation}
\rho(E) = \frac{1}{\pi} \lim_{\sigma \rightarrow 0} \lim_{\Lambda
  \rightarrow \infty} \EE\left[\textrm{Im } G(E+i\sigma; 0, 0) \right]
\end{equation}
where $G$ is the resolvent, or Green's function, of the system 
\begin{equation}
G(z) = (H-z)^{-1}
\end{equation}
The limit $\Lambda \rightarrow \infty$ stands for the fact that we must work
in a finite volume to have well defined quantities and then take the 
thermodynamic limit. 

Thus the problem amounts to studying the mean Green's function. Perturbations
suggest that 
\begin{equation}
\EE\left[G(E+i\sigma)\right] \sim \frac{1}{p^{2}-E -i\sigma -\Sigma}
\end{equation}
where the self-energy $\Sigma$ is given at leading order by the Born 
approximation, \cite{Weg,OW}  
\begin{eqnarray}
C &=& \frac{1}{p^{2}-E -i\sigma -\Sigma_{Born}} \\
\Sigma_{Born} &=& \lambda^{2} \, \EE(V C V) 
\end{eqnarray}
which yields a finite imaginary part of order $\lambda^{2}$. 

In this paper, we derive a Ward type identity which should allow to control
the mean Green's function for \emph{initial} imaginary part of order
$\lambda^{2+\varepsilon}$, \emph{i.e.} much smaller than the expected final
imaginary part. This is not enough to go to the limit $\sigma \rightarrow 0$
which is in fact equivalent to $\sigma \simeq \lambda^{3}$ thanks to spectral
averaging techniques \cite{CH} or equivalently complex translation of the
potential \cite{Poi1}. Nevertheless, we think that this kind of identity
should play a role in studying the mean Green's function of the model and in
proving its expected long range decay. 

We give first a heuristic presentation of this identity which is a little
bit complicated. Then we will derive it in a simplified model in a single cube
of size $\lambda^{-2-\varepsilon}$. This result, when combined with a polymer 
expansion of the resolvent would allow to control the thermodynamic limit of
the model with $\sigma=\lambda^{2+\varepsilon}$, in the same way than 
(\cite{Poi1,Poi2}).


\section{Phase space picture and matrix model}
Our study is based on a phase space multiscale analysis \cite{Poi1,MPR,Poi2}. 
We divide the momentum space into slices such that in the $j^{th}$ slice
$\Sigma_{j}$, we have $M^{-j-1}\leqslant |p^{2}-E| \leqslant M^{-j}$ for some
integer $M \geqslant 2$. Then the real space is divided into lattices
$\DD_{j}$ of cubes of dual size $M^{j}$. 

The point is that the potential $V$ seen as an operator $\mathbf{V}$ has a 
kernel in momentum space given by 
\begin{equation}
\mathbf{V}(p, q) = \hat{V}(p-q). 
\end{equation}
When $p$ and $q$ are restricted to low slices, \emph{i.e.} very close to
$p^{2}=E$, knowing the momentum transfer $p-q$ allows to recover back the pair
$\{p, -q\}$ \cite{FMRT1,FMRT2} so that the potential has a very 
strong matrix flavor \cite{MPR}. 

We call $\eta_{j}$ a smoothed projector on the slice $\Sigma_{j}$ that we
further divide into angular sectors $S_{\alpha}^{j}$ of width $M^{-j/2}$, 
corresponding to some $\eta_{\alpha_{j}}$. We write $\bar{\alpha}$ for the 
opposite sector to $\alpha$ and we introduce also the notation 
\begin{equation}
\eta_{\bar{j}} = \sum_{k \geqslant j} \eta_{k}
\end{equation}

For any operator $A$ we write 
\begin{equation}
A^{jk} = \eta_{j} A \eta_{k}
\end{equation}

Finally, for any lattice $\DD_{j}$ of cubes $\Delta$, we can make an
orthogonal decomposition of the field $V$ into a sum of fields
$V_{\Delta}$, the support of $V_{\Delta}$ being on a close neighborhood of
$\Delta$ \cite{Poi1}. Then, using the matrix aspect of the potential, we can
derive the following estimates \cite{Poi1,Poi2}

\begin{lemma}
\ 
\label{lemprob1}

There are constants $K_{1}$ and $K_{2}$ such that for all
$j\leqslant k$, $a\geqslant 1$ and $\Delta \in \DD_{k}$ 
\begin{equation}
\PP\left(\|V_{\Delta}^{\bar{j} \bar{k}}\| \geqslant a K_{1} M^{-j/2}\right)
  \leqslant K_{2} \, e^{- a^{2} M^{\frac{k}{2} - \frac{j}{3}}} 
\end{equation}
\end{lemma}

\begin{lemma} 
(Tadpole-free operators) 
\label{lemprob2}

There are constants $K_{1}$ and $K_{2}$ such that for all
$j\leqslant k$, $a\geqslant 1$ and $\Delta, \Delta' \in \DD_{k}$ 
\begin{equation}
\PP\left(\|:V_{\Delta}^{\bar{k} j} C V_{\Delta'}^{j \bar{k}}:\| 
  \geqslant a K_{1} M^{-\frac{(k-j)}{2}}\right)
  \leqslant K_{2} \, e^{- a^{2} M^{\frac{j}{6}}} 
\end{equation}
where $::$ stands for the Wick ordering 
\begin{equation}
:V_{\Delta}^{\bar{k} j} C V_{\Delta'}^{j \bar{k}}: = 
  V_{\Delta}^{\bar{k} j} C V_{\Delta'}^{j \bar{k}} 
  - \left<V_{\Delta}^{\bar{k} j} C V_{\Delta'}^{j \bar{k}}\right>
\end{equation}
\end{lemma}

\begin{lemma}
(Almost diagonal operators)
\label{lemprob3}

There are constants $K_{1}$ and $K_{2}$ such that for all 
$j \leqslant k$, $0<r<1$, $a\geqslant 1$ and $\Delta \in \DD_{k}$ 
\begin{equation}
\PP\left(\| \ \!\!^{(r)} \lbar{V}_{\Delta}^{\bar{j} \bar{k}}\| \geqslant 
  a K_{1} M^{-\frac{j}{2}(1+\frac{r}{2})}\right) 
  \leqslant K_{2} \, e^{-a^{2} M^{\frac{k}{2}-\frac{j}{3}-\frac{rj}{6}}} 
\end{equation}
where
\begin{equation}
^{(r)} V_{\Delta}^{jk} \equiv \sum_{|\alpha-\beta| \leqslant M^{-rj/2}}
\eta_{\alpha_{j}} V_{\Delta} \eta_{\beta_{k}} + \sum_{|\alpha-\bar{\beta}|
  \leqslant M^{-rj/2}} \eta_{\alpha_{j}} V_{\Delta} \eta_{\beta_{k}}
\end{equation} 
\end{lemma}


\section{Result}
We are looking at a simplified model in a single cube: in 
$\RR^{2}/_{\textstyle \lambda^{-2-\varepsilon} \ZZ^{2}}$ we consider the 
following Hamiltonian 
\begin{equation}
H = -\Delta + \lambda V
\end{equation}
where $V$ is a Gaussian random field with translation invariant covariance 
$\xi \in \Ccal^{\infty}_{0}(\RR^{2})$ such that $\xi^{1/2} \in
\Ccal^{\infty}_{0}(\RR^{2})$ (the square root being taken in operator sense). 
The result could be extended to $\xi \in \mathcal{S}(\RR^{2})$ but the
development would be much heavier.  

We are interested in computing the mean Green's function 
\begin{equation}
\bar{G}(E- i \sigma) = \int \!\! d\mu(V) \, (H-E-i \sigma)^{-1}
\end{equation}
for which we have the following result 

\begin{theorem}
(Ward type identity)
\label{thward}

There exists $\nu >0$ such that for all $\sigma \leqslant \lambda^{2}$ 
\begin{equation}
\label{eqtheoremi}
\| \bar{G}(E) - C \|\leqslant O(1)
\left(\frac{\lambda^{2+\nu}}{\sigma}\right)^{3} \frac{1}{\sigma} 
\end{equation}
where $C$ is the renormalized propagator at leading order (Born
approximation). 
\end{theorem}

Moreover, we can iterate the development in order to obtain an asymptotic
expansion to all orders. 

This result together with a polymer expansion \cite{Poi1} would allow to
control the thermodynamic limit for imaginary part 
\begin{equation}
\sigma \geqslant \lambda^{2+\varepsilon}. 
\end{equation}
Therefore we can investigate the mean Green's function up to a region with
imaginary part much smaller than the expected final one. 
We cannot for the moment go to the real axis which would require to have 
$\nu>1$. Nevertheless, we think
that this kind of identity should play a role in the study of the mean Green's
function and of the density of states.


\section{Heuristic presentation}
\label{secheuristic}
Let us assume that we are working in the region of momenta $|p^{2}-E|\leqslant
M^{-j_{1}}$, where 
\begin{eqnarray}
j_{1} &=& (1-\nu_{1}) j_{0} \\ 
\lambda^{2} (\log \lambda^{-1})^{2} &\simeq& M^{-j_{0}}
\end{eqnarray}

We are looking at the off-diagonal part of the potential, \emph{i.e.} the 
$V_{\alpha \beta}$'s such that $\alpha$ is quite different from $\beta$
or $\bar{\beta}\equiv \pi + \beta$. We know that in this case, sectors are
preserved up to $M^{-j_{0}/2}$ and even better \cite{Poi2}. 
If we introduce a counter-term $\delta$ 
and perform one step of perturbation, \emph{e.g.} by putting an
interpolation parameter on the potential and the
counter-term, we get a remainder term which looks like 

\vspace{0.5em}
\begin{equation}
R = - \left< 
\begin{picture}(34,15)
\put(0,0){\psfig{figure=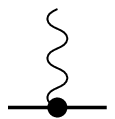}}
\end{picture}
\right>
- \left< 
\begin{picture}(32,15)
\put(0,0){\psfig{figure=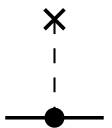}}
\put(20,13){$\scriptstyle \delta$}
\end{picture}
\right>
\end{equation}

We can integrate the $V$ by parts and get 

\vspace{0.5em}
\begin{equation}
R = 2 \left< 
\begin{picture}(75,15)
\put(0,0){\psfig{figure=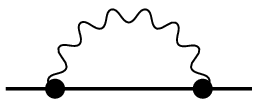}}
\end{picture}
\right>
- \left< 
\begin{picture}(32,15)
\put(0,0){\psfig{figure=c_term.eps}}
\end{picture}
\right>
\end{equation}

Sector conservation tells us that there are two possible configurations: 
\begin{eqnarray}
\begin{picture}(75,35)
\put(0,0){\psfig{figure=order2.eps}} 
\put(8,-7){$\scriptstyle \alpha$}
\put(20,-7){$\scriptstyle \beta$}
\put(49,-7){$\scriptstyle \bar{\alpha}$}
\put(61,-7){$\scriptstyle \bar{\beta}$}
\end{picture}
&\equiv& 
\begin{picture}(75,35)
\put(0,0){\psfig{figure=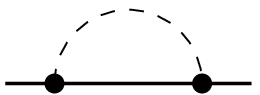}} 
\put(8,-7){$\scriptstyle \alpha$}
\put(20,-7){$\scriptstyle \alpha$}
\put(49,-7){$\scriptstyle \bar{\beta}$}
\put(61,-7){$\scriptstyle \bar{\beta}$}
\end{picture} \label{eqabab}
\\
\begin{picture}(75,30)
\put(0,0){\psfig{figure=order2.eps}} 
\put(8,-7){$\scriptstyle \alpha$}
\put(20,-7){$\scriptstyle \beta$}
\put(49,-7){$\scriptstyle \beta$}
\put(61,-7){$\scriptstyle \alpha$}
\end{picture}
&\equiv& 
\begin{picture}(50,60)
\put(0,0){\psfig{figure=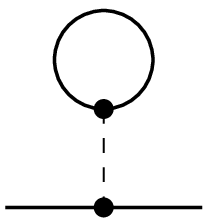,height=50pt}} 
\put(16,-7){$\scriptstyle \alpha$}
\put(28,-7){$\scriptstyle \alpha$}
\put(16,17){$\scriptstyle \beta$}
\put(28,17){$\scriptstyle \beta$}
\end{picture}
\label{tadgraph}
\end{eqnarray}

We used the fact that $V$ being almost ultra-local we can identify both ends
of its propagator (the wavy line) and replace it by the dashed line which
corresponds to the low momentum channel. Furthermore, for the first term, we
used 
\begin{equation}
(\eta_{\beta} G \eta_{\bar{\alpha}})(x, x) = (\eta_{\beta} G
\eta_{\bar{\alpha}})^{\mathrm{t}}(x, x) = (\eta_{\alpha} G
\eta_{\bar{\beta}})(x, x)  
\label{eqretournement}
\end{equation}
where $A^{\mathrm{t}}$ stands for the transposed operator, whose kernel is
\begin{equation}
A^{\mathrm{t}} (x, y) = A (y, x)
\end{equation}

The first configuration corresponds to the insertion of two fields of very low
momentum, \emph{i.e.} to almost diagonal operators, and as such is very
small. Thus we just need to see how the second term, which is the insertion of
a tadpole, will kill the counter-term, at least at leading order. 

First, we can remark that if $\alpha$ and $\beta$ are far enough from each
other, the momentum flowing into the loop has a size $M^{-j_{1}}$, being
at the intersection of two tubes of size $M^{-j_{0}/2} \times M^{-j_{1}}$. But
we can go further: if all the incoming legs at the vertex are in the ``very 
low'' slice $\Sigma_{j_{0}}$, the ingoing momentum has a size $M^{-j_{0}}$. 
This means that either we have a very small momentum flowing into the loop or
one of the four legs is ``high'', which means that $|p^{2}-E|$ is large. This
allows to earn a small factor. 

If we set the counter-term equal to the tadpole with the bare propagator $C_{0}$, we are led to study 

\vspace{1.5em} 
\begin{equation}
\Gamma_{\alpha \beta}(k) = \left<
\begin{picture}(50,20)
\put(0,0){\psfig{figure=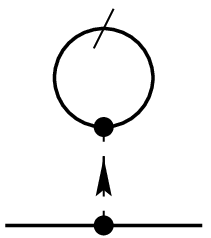,height=50pt}} 
\put(16,-8){$\scriptstyle \alpha$}
\put(28,-8){$\scriptstyle \alpha$}
\put(12,19){$\scriptstyle \beta$}
\put(30,19){$\scriptstyle \beta$}
\put(15,8){$\scriptstyle k$}
\end{picture}
\right>
\end{equation}
where the slashed line stands for $G-C_{0}$. 
In momentum space the contribution of the loop is 
\begin{equation}
\int \!\! dp \, \eta_{\beta}(p) \eta_{\beta}(p+k) (G-C_{0})(p+k, p). 
\end{equation}
The key point is then to notice that when $p$ is close to $k_{\beta}$, the
center of the sector $\beta$ in momentum space, one can write 
\begin{eqnarray}
2 k.k_{\beta} + k^{2} &=& \left[(p+k)^{2} -E -i \sigma \right] - 
\left[p^{2}-E-i \sigma \right] - 2 k.(p-k_{\beta}) \\
&=& C_{0}^{-1}(p+k) - C_{0}^{-1}(p) + O\left(|k| |p-k_{\beta}|\right)
\end{eqnarray}
Using the resolvent identity 
\begin{equation}
G-C_{0} = - C_{0} \lambda V G = - G \lambda V C_{0}
\end{equation}
we get 
\vspace{2em}
\begin{equation}
(2 k.k_{\beta} + k^{2}) \Gamma_{\alpha \beta}(k) \simeq 
\left<
\begin{picture}(50,20)
\put(0,0){\psfig{figure=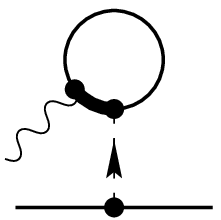,height=50pt}} 
\put(17,-7){$\scriptstyle \alpha$}
\put(29,-7){$\scriptstyle \alpha$}
\put(15,16){$\scriptstyle \beta$}
\put(30,16){$\scriptstyle \beta$}
\end{picture}
-
\begin{picture}(50,20)
\put(0,0){\psfig{figure=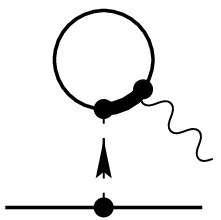,height=50pt}} 
\put(17,-7){$\scriptstyle \alpha$}
\put(29,-7){$\scriptstyle \alpha$}
\put(12,16){$\scriptstyle \beta$}
\put(28,16){$\scriptstyle \beta$}
\end{picture}
\right>
\label{eqdiagJRJL}
\end{equation}
where the thick line stands for $\eta_{\beta}$. 

The perturbative reformulation of this identity is the following. Starting
from a vertex function 
\begin{equation}
\Gamma_{2, 1}^{(\beta)} (p, k) = C_{\beta}(p) C_{\beta}(p+k)
\end{equation}
we have the identity 
\begin{equation}
(2 k.k_{\beta} + k^{2}) \Gamma_{(2, 1)}^{(\beta)} (p, k) \simeq 
C_{\beta}(p) - C_{\beta}(p+k) 
\end{equation}
This looks very much like a perturbative Ward identity in Quantum 
Electrodynamics where we express the vertex 
function $\Gamma_{2, 1}$ (between an electron, a positron and a photon) 
in terms of the difference of two electron propagators.  

Once again, we integrate by parts the $V$ which has been taken down. This leads
to six possible graphs 
\begin{displaymath}
\begin{picture}(50,50)
\put(0,0){\psfig{figure=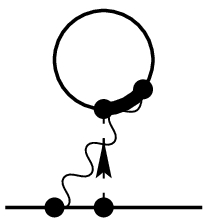,height=50pt}} 
\end{picture}
\quad
\begin{picture}(50,50)
\put(0,0){\psfig{figure=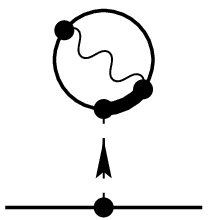,height=50pt}} 
\end{picture}
\quad
\begin{picture}(50,50)
\put(0,0){\psfig{figure=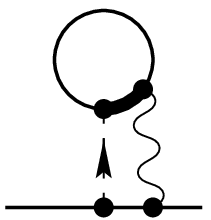,height=50pt}} 
\end{picture}
\quad 
\begin{picture}(50,50)
\put(0,0){\psfig{figure=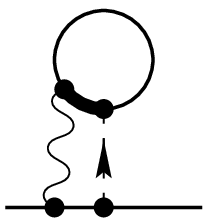,height=50pt}} 
\end{picture}
\quad
\begin{picture}(50,50)
\put(0,0){\psfig{figure=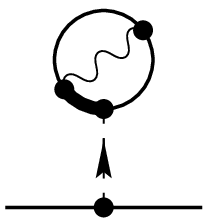,height=50pt}} 
\end{picture}
\quad
\begin{picture}(50,50)
\put(0,0){\psfig{figure=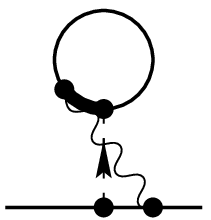,height=50pt}} 
\end{picture}
\end{displaymath}

Then we use sector conservation and perform various unfolding operations. 
In order to illustrate the process, let us show how it works on the following
typical term (using the fact that a wavy line is almost a $\delta$ function)

\begin{equation}
\begin{picture}(60,50)
\put(0,0){\psfig{figure=graph_C56.eps,height=60pt}} 
\put(30,40){$\scriptstyle \beta$}
\put(47,-7){$\scriptstyle \beta$}
\end{picture} 
= 
\begin{picture}(95,50)
\put(0,0){\psfig{figure=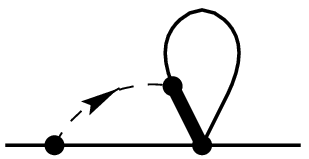}}
\put(42,10){$\scriptstyle \beta$}
\put(70,-7){$\scriptstyle \beta$}
\end{picture}
=
\raisebox{-5pt}{%
\begin{picture}(100,50)
\put(0,0){\psfig{figure=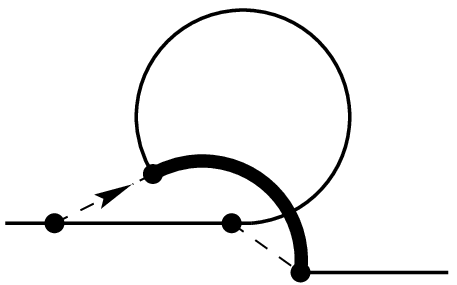,height=60pt}}
\put(50,30){$\scriptstyle \beta$}
\put(80,-8){$\scriptstyle \beta$}
\end{picture}}
=
\raisebox{-19pt}{%
\begin{picture}(80,50)
\put(0,0){\psfig{figure=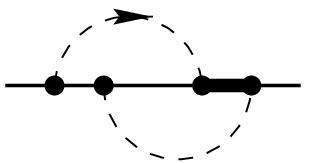}}
\put(62,30){$\scriptstyle \beta$}
\put(78,12){$\scriptstyle \beta$}
\end{picture}}
\end{equation}

In the end we obtain 12 terms which are 
\begin{eqnarray}
\hspace{-2em}
(2 k.k_{\beta} + k^{2}) \Gamma_{\alpha \beta}(k) &\simeq& 
\left<
\begin{picture}(76,20)
\put(0,-14){\psfig{figure=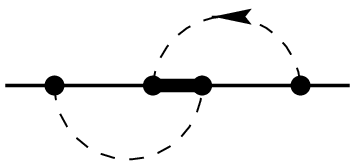,width=75pt}} 
\put(9,8){$\scriptstyle \gamma$}
\put(33,-8){$\scriptstyle \beta$}
\put(60,-8){$\scriptstyle \alpha$}
\end{picture}
+
\begin{picture}(76,20)
\put(0,-17){\psfig{figure=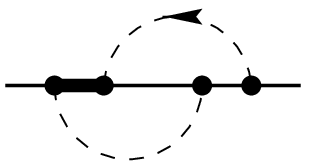,width=75pt}} 
\put(43,8){$\scriptstyle \gamma$}
\put(17,-8){$\scriptstyle \bar{\beta}$}
\put(60,-8){$\scriptstyle \alpha$}
\end{picture}
+
\begin{picture}(35,60)
\put(0,0){\psfig{figure=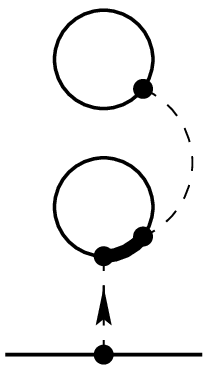,height=60pt}} 
\put(22,11){$\scriptstyle \beta$}
\end{picture}
\right. \nonumber \\ 
&& \left. + \,
\begin{picture}(41,50)
\put(0,0){\psfig{figure=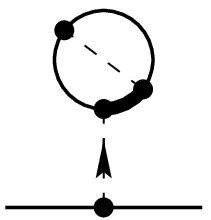,height=45pt}} 
\put(26,16){$\scriptstyle \beta$}
\end{picture}
+ 
\begin{picture}(76,20)
\put(0,-17){\psfig{figure=graph_C5.eps,width=75pt}} 
\put(54,8){$\scriptstyle \beta$}
\end{picture}
+
\begin{picture}(76,20)
\put(0,-14){\psfig{figure=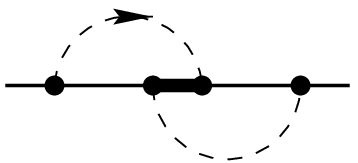,width=75pt}} 
\put(34,-8){$\scriptstyle \bar{\beta}$}
\end{picture}
\right. \nonumber \\
&& \left. - \,
\begin{picture}(76,20)
\put(0,-17){\psfig{figure=graph_C2.eps,width=75pt}} 
\put(17,-8){$\scriptstyle \beta$}
\end{picture}
- 
\begin{picture}(76,20)
\put(0,-14){\psfig{figure=graph_C1.eps,width=75pt}} 
\put(33,-8){$\scriptstyle \bar{\beta}$}
\end{picture}
- 
\begin{picture}(35,75)
\put(0,0){\psfig{figure=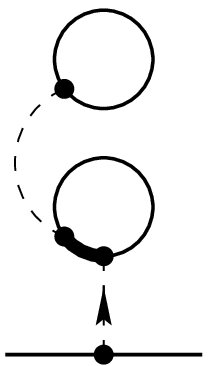,height=60pt}} 
\put(8,11){$\scriptstyle \beta$}
\end{picture}
\right. \nonumber \\ 
&&\left. - \,
\begin{picture}(41,65)
\put(0,0){\psfig{figure=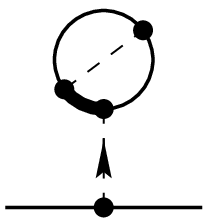,height=45pt}} 
\put(12,16){$\scriptstyle \beta$}
\end{picture}
-
\begin{picture}(76,20)
\put(0,-14){\psfig{figure=graph_C6.eps,width=75pt}} 
\put(36,-8){$\scriptstyle \beta$}
\end{picture}
-
\begin{picture}(76,20)
\put(0,-17){\psfig{figure=graph_C5.eps,width=75pt}} 
\put(54,8){$\scriptstyle \bar{\beta}$}
\end{picture}
\right>
\label{termesward}
\end{eqnarray}
\goodbreak

We have two types of graphs: 
\begin{itemize}
\item graphs without electron loops, containing very low momentum field 
  insertions, 
\item and graphs with remaining electron loops. 
\end{itemize}
The graphs in the first category are small. In the second category, the 3rd, 
4th, 9th and 10th graphs form two pairs which almost compensate each other 
($3-9$ and $4-10$). 

The point is that when all the incoming legs at the various vertices are
``very low'', the momenta flowing into the loops have size $M^{-j_{0}}$. This
implies that the dashed lines stand for \emph{propagators} which decay on a 
length scale $M^{j_{0}}$. But $\eta_{\beta}$ (the thick line) decays on a
scale $M^{j_{1}} \ll M^{j_{0}}$, therefore it is almost-local with respect
to the scale $M^{j_{0}}$ and we can approximate it by a point. This means 
that  the graphs number $3$ and $9$ as well as the number $4$ and $10$ 
compensate each other, up to a gradient term in $M^{-(j_{0}-j_{1})}$.  

This conclude our heuristic description of the reasons for which Theorem
\ref{thward} holds. The next section is devoted to the proof. 


\section{Proof of Theorem \protect \ref{thward}}
\subsection{Notations}
First of all, we introduce various notations 
\begin{itemize}
\item We define $\ \!\! _{\downarrow} \equiv \eta_{\downarrow} = 
\eta_{\bar{j}_{1}}$ et $\ \!\! _{\uparrow} \equiv \eta_{\uparrow} = 
(1-\eta_{\bar{j}_{1}})$, so that every operator $A$ can be written 
\begin{equation}
A = \ \!\!_{\uparrow} A_{\uparrow} + \ \!\!_{\uparrow} A_{\downarrow} + \
\!\!_{\downarrow} A_{\uparrow} + \ \!\!_{\downarrow} A_{\downarrow} 
\end{equation} 
\item for $0<r<1$ to be fixed later, we define 
\begin{equation}
\label{defsim}
\alpha \sim \beta \Leftrightarrow |\alpha - \beta| \leqslant
M^{-r \frac{j_{0}}{2}} 
\end{equation} 
\item for $K=O(1)$ we note 
\begin{equation}
\label{defsimeq}
\alpha \simeq \beta \Leftrightarrow |\alpha - \beta| \leqslant K
M^{-j_{0}/2} 
\end{equation} 
\item finally we introduce 
\begin{equation}
j_{2} = (1-\nu_{2}) j_{0} > j_{1}
\end{equation}
\end{itemize}


\subsection{Starting the development}
We define the counter-term through the self-consistent equation 
\begin{eqnarray}
\Sigma (x-y) &=& \lambda^{2} \xi (x-y) C(x-y) \\[5pt] 
C &=& \frac{1}{p^{2}-E-i\sigma -\Sigma}
\end{eqnarray} 
Then we set 
\begin{eqnarray}
G(t) &=& \frac{1}{p^{2}-E-i\sigma -\Sigma +t\lambda V +t^{2}
  \Sigma} \\[5pt]
\bar{G}(t) &=& \EE \left[G(t) \right] 
\end{eqnarray}
so that we can write 
\begin{equation}
\bar{G} = \bar{G}(1) = \bar{G}(0) + \int_{0}^{1} 
\partial_{t} \bar{G}(t) \, dt 
= C + \int_{0}^{1} \partial_{t} \bar{G}(t) \, dt
\end{equation}

Then our problem reduces to the study of 
\begin{eqnarray}
\partial_{t}\bar{G} &=& - \left< G (\lambda V +2t \Sigma) G \right>
  \label{befipp} \\
  &=& 2t \int \!\! dx \, dy \, \left< G(., x) \left[\lambda^{2}
  \xi(x, y) G(x, y) - \Sigma(x, y)\right] G(y, .) \right> \label{aftipp} \\
  &=& 2t \int \!\! dx \, dy \, \left< G(., x) \left[\lambda^{2}
  \xi(x, y) (G-C) (x, y) \right] G(y, .) \right> \\
  &\equiv& 2t \left< G \left[\lambda^{2} \xi * (G-C) \right] G 
  \right>
\end{eqnarray}
where we have integrated the $V$ by parts from (\ref{befipp}) to 
(\ref{aftipp}). 

If we plug in the ``high'' and ``low'' slices we get 
\begin{eqnarray}
\partial_{t}\bar{G} &=& \sum_{i_{1}, \ldots, i_{4} \in 
  \{ \downarrow, \uparrow\}} 2t \left< G \eta_{i_{1}} \left[ \lambda^{2} 
    \xi * \eta_{i_{2}} (G - C) \eta_{i_{3}} \right]  
  \eta_{i_{4}} G \right> \\ 
&=& \sum_{i_{1}, \ldots, i_{4} \in 
  \{ \downarrow, \uparrow\}} \partial_{t}\bar{G}_{i_{1} \ldots i_{4}}
\end{eqnarray}


\subsection{Higher part}
We can quite easily deal with the case $(i_{1}, \ldots, i_{4}) \neq 
(\downarrow, \ldots, \downarrow)$ because we have a high leg. 

\begin{lemma}
\label{lemhighpart}
\begin{equation}
\sum_{(i_{1}, \ldots, i_{4}) \neq (\downarrow, \ldots, \downarrow)} 
  \| \partial_{t} \bar{G}_{i_{1} \ldots i_{4}}\| \leqslant O(1)
  \lambda^{\nu_{1}} \left(\frac{\lambda^{2}}{\sigma}\right)^{2}
  \times \frac{1}{\sigma}
\end{equation}
\end{lemma}
\begin{proof}
We start with 
\begin{equation}
G(t) - C = - C (\lambda V + t^{2} \Sigma) G(t)
\end{equation}
then, we write the covariance $\xi$ as the integration of two insertions of an
auxiliary field $U$.  
\begin{equation}
\partial_{t} \bar{G}_{i_{1} \ldots i_{4}} = -2t \int \!\! 
  G \eta_{i_{1}} (\lambda U) \eta_{i_{2}} C 
  (\lambda V + t^{2} \Sigma) G \eta_{i_{3}}
  (\lambda U) \eta_{i_{4}} G \, d\mu(V) \, d\mu(U)  
\end{equation}

At this point we perform a large field versus small field decomposition
\cite{Poi1,Poi2} 
\begin{equation}
1 = \varepsilon (U, V) + (1-\varepsilon) (U, V)
\end{equation}
where $\varepsilon$ is a smooth function which forces 
\begin{eqnarray}
\| V^{\bar{j} \bar{k}}\| &\leqslant& O(1) M^{-\frac{j_{1}}{2}}
  M^{\tau_{1}\max (\frac{j_{1}-k}{2}, 0)} \\ 
\| U^{\bar{j} \bar{k}}\| &\leqslant& O(1) M^{-\frac{j_{1}}{2}}
  M^{\tau_{1}\max(\frac{j_{1}-k}{2}, 0)} \\ 
\| U^{\bar{k} j} C^{j} V^{j \bar{k}} \| &\leqslant& O(1)
  M^{-(\frac{k-j}{2})} M^{\tau_{1}\max[(j_{1}-j), 0]} 
\end{eqnarray}
The last condition is possible because $U$ and $V$ cannot contract together,
therefore we can use lemma \ref{lemprob2} on tadpole-free
operators. Thanks to lemmas \ref{lemprob1} and \ref{lemprob2}, 
we find that the large field contribution will be small, of order 
\begin{equation}
\EE\left[\| \partial_{t} G_{i_{1} \ldots i_{4}} \| (1-\varepsilon) \right] 
  \leqslant O(1) \, \lambda^{-2} \, 
  e^{-\kappa \lambda^{-2 \tau_{1}(1-\nu_{1})}} \times \frac{1}{\sigma^{3}}, 
\end{equation}
for some constant $\kappa$. 

In the small field region, we will use the fact that we have a high leg so
that in some sense we can make perturbations. Suppose that $\eta_{i_{4}}$
is the high leg at scale $k_{0}$, the operator $\lambda U^{i_{3} i_{4}}$ will
have a size $\lambda M^{-k_{0}/2} \gg \lambda^{2}$. But if we perform a 
resolvent expansion on $\lambda U^{i_{3} i_{4}} G$ we will get $\lambda
U^{i_{3} i_{4}} C^{k_{0}} \lambda V G$ thus we earn an extra factor $\lambda
U^{i_{3} i_{4}} C^{k_{0}}$ whose norm is $\lambda M^{k_{0}/2} \ll 1$. Then we
can iterate the process until either we fall back to a scale of order $j_{0}$
or we have earned enough small factors. 

\begin{lemma}
  (stairway expansion)
\begin{eqnarray}
(1-\eta_{\bar{m}}) G &=& \sum_{k_{0}<m} C^{k_{0}} \left[ 
  1 - (t^{2} \Sigma + \lambda V \eta_{\bar{j_{0}}}) G \right] 
  \nonumber \\
&& \hspace{-1cm} - \sum_{{k_{0}<m} \atop {k_{1}<j_{0}}} C^{k_{0}} 
  \lambda V C^{k_{1}}
  \left[ 1 - (t^{2} \Sigma + \lambda V \eta_{\bar{k_{1}}}) G
  \right] \nonumber \\
&& \hspace{-1cm} + \sum_{
  \begin{array}{c}
    \vspace{-20pt} \\
    \scriptstyle k_{0}<m \\[-6pt]
    \scriptstyle 0<k_{1}<j_{0} \\[-6pt]
    \scriptstyle k_{2}<k_{1} \\[-10pt]
  \end{array}} 
  C^{k_{0}} \lambda V C^{k_{1}} \lambda V C^{k_{2}}
  \left[1 - (t^{2} \Sigma + \lambda V \eta_{\bar{k_{2}}}) G 
  \right] - \ldots 
  \label{eqstair}
\end{eqnarray}
where $C^{k} = \eta_{k} C$ and $\eta_{\bar{0}}=1$.

Of course, we have a similar result for $G (1-\eta_{\bar{m}})$ by expanding
to the left. 
\end{lemma}

\begin{proof}
The proof is by induction on $m$ thanks to the resolvent identity 
\begin{equation}
G = C - C (t^{2} \Sigma + \lambda V) G 
\end{equation}
that we write as 
\begin{equation}
\eta_{p} G = C^{p} \left[1 - C (t^{2} \Sigma + \lambda V
  \eta_{\bar{p}}) G \right] - \sum_{q<p} C^{p} \lambda V \eta_{q} G
\end{equation}
\end{proof}

In equation (\ref{eqstair}) we can group the various terms according to
whether they end by a $C$, a $\Sigma G$ or a $V G$. Then we can introduce a
diagrammatic representation 
\begin{eqnarray}
(1-\eta_{\bar{j}_{0}}) G &=& 
\begin{picture}(35,10)
\put(0,-2){\psfig{figure=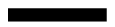}} 
\put(8,10){$\scriptstyle k_{0} < j_{0}$}
\end{picture} = A_{V} + A_{\Sigma} + A_{C} \\ 
A_{V} &=& \raisebox{-1.1cm}{%
\begin{picture}(35,10)
\put(0,0){\psfig{figure=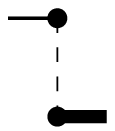}} 
\put(5,41){$\scriptstyle k_{0}$}
\put(20,12){$\scriptstyle j_{0}$}
\end{picture}}
+ 
\psfig{figure=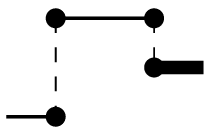} + \ldots + 
\psfig{figure=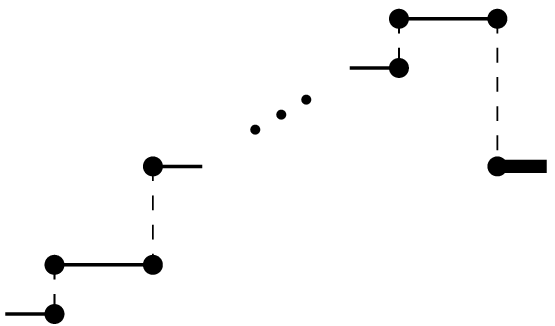}
\end{eqnarray}
$A_{\Sigma}$ and $A_{C}$ can be represented in the same way by changing the
rightmost term. 

We apply the stairway expansion on each leg which has its momentum above the 
scale $j_{0}$ and is not linked to a long enough stairway. This allows to show
that each field insertion behaves like $O(1) \lambda^{2}$. Furthermore, we
earn a factor $\lambda M^{j_{1}/2} M^{\tau_{1} j_{1}/2} \sim
\lambda^{\nu_{1}-\tau_{1}}$, thanks to the $\lambda U$ which have a high leg. 
Indeed, the corresponding insertion of $\lambda U$ will transform into the
insertion of a sum of stairways. The $A_{V}$ part has the following form:  
\begin{itemize}
\item the second order insertion is 
\begin{equation}
A_{2} \equiv \sum_{k_{0}<j_{1}}
\begin{picture}(50,47)
\put(0,0){\psfig{figure=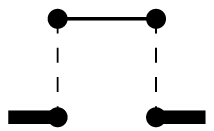}}
\put(7,-6){$\scriptstyle j_{0}$}
\put(50,-6){$\scriptstyle j_{0}$}
\put(0,15){$\lambda U$}
\put(50,15){$\lambda V$}
\put(28,40){$\scriptstyle k_{0}$}
\end{picture} 
\end{equation}
Since $U$ and $V$ cannot contract together, the small field condition tells us
that 
\begin{equation}
\| A_{2} \| \leqslant O(1) \sum_{k_{0}<j_{1}} \lambda^{2}
M^{-(\frac{j_{0}-k_{0}}{2})} \leqslant O(1) \lambda^{2}
M^{-(\frac{j_{0}-j_{1}}{2})} 
\end{equation}
\item at third order, we have two possible configurations 
\begin{equation}
A_{3}^{(1)} \equiv \sum_{{k_{0}<j_{1}} \atop {k_{1} \leqslant k_{0}}}
\begin{picture}(90,60)
\put(0,0){\psfig{figure=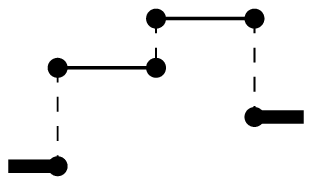}}
\put(7,-6){$\scriptstyle j_{0}$}
\put(0,15){$\lambda U$}
\put(25,38){$\lambda V$}
\put(77,28){$\lambda V$}
\put(29,24){$\scriptstyle k_{0}$}
\put(57,38){$\scriptstyle k_{1}$}
\put(77,8){$\scriptstyle > j_{1}$}
\end{picture} 
\quad \textrm{and} \quad 
A_{3}^{(2)} \equiv \sum_{{k_{0}<j_{1}} \atop {k_{1} > k_{0}}}
\begin{picture}(90,60)
\put(0,0){\psfig{figure=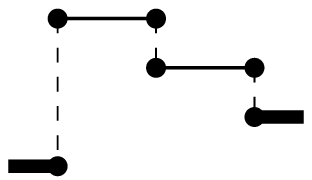}}
\put(7,-6){$\scriptstyle j_{0}$}
\put(0,22){$\lambda U$}
\put(54,40){$\lambda V$}
\put(79,23){$\lambda V$}
\put(29,54){$\scriptstyle k_{0}$}
\put(57,23){$\scriptstyle k_{1}$}
\end{picture} 
\end{equation}
For $A_{3}^{(1)}$, one finds 
\begin{eqnarray}
\hspace{-1em} 
\|A_{3}^{(1)} \| &\lesssim & \lambda M^{-k_{0}/2} M^{k_{0}} 
  \lambda M^{-k_{1}/2} M^{\tau_{1}(\frac{j_{1}-k_{0}}{2})} M^{k_{1}} 
  \lambda M^{-k_{1}/2} \\
  &\lesssim & \lambda^{2} \left(\lambda M^{j_{1}/2} \right) 
    M^{-(1-\tau_{1})(\frac{j_{1}-k_{0}}{2})} \\ 
  &\lesssim & \lambda^{2} \left(\lambda M^{j_{1}/2} \right) 
\end{eqnarray}
In the same way, one finds for $A_{3}^{(2)}$ 
\begin{eqnarray}
\hspace{-1em} 
\|A_{3}^{(2)} \| &\lesssim & \lambda^{2} M^{-(\frac{k_{1} -k_{0}}{2})} 
  M^{\tau_{1}(j_{1}-k_{0})} M^{k_{1}} \lambda M^{-k_{1}/2} 
  M^{\tau_{1}(\frac{j_{1}-k_{1}}{2})} \\
  &\lesssim & \lambda^{2} \left(\lambda M^{j_{1}/2} \right)  
    M^{-(1- 3 \tau_{1})(\frac{j_{1}-k_{0}}{2})} \\ 
  &\lesssim & \lambda^{2} \left(\lambda M^{j_{1}/2} \right) 
\end{eqnarray}
\item fourth and higher order terms can be treated similarly. 
  We find the same power counting with more and more small
  factors as the order increases. 
\end{itemize}

We do the same to bound the $A_{C}$ and $A_{\Sigma}$ parts and get the 
announced result 
\begin{equation}
\|\partial_{t} \bar{G}_{i_{1} \ldots i_{4}} \| \leqslant O(1)
  \lambda^{\nu_{1}} \left(\frac{\lambda^{2}}{\sigma}\right)^{2}
  \times \frac{1}{\sigma} 
\end{equation}
\end{proof}


\subsection{Lower part} 
We introduce the angular sectors
\begin{equation}
\label{defGbas}
\partial_{t} \bar{G}_{\downarrow} \equiv  \partial_{t} 
  \bar{G}_{\downarrow \ldots
  \downarrow} = 2t \sum_{\alpha_{1} \ldots \alpha_{4}} 
  \left<G \eta_{\alpha_{1}} \left[\lambda^{2} \xi * \eta_{\alpha_{2}}
  (G-C) \eta_{\alpha_{3}} \right] \eta_{\alpha_{4}} G\right>
\end{equation}

First we extract the \emph{degenerate} part of the sum over the
sectors, \emph{i.e.} the part $(\alpha_{1} \sim \alpha_{2},
\bar{\alpha}_{2})$. In order to do so, we define the almost diagonal part of 
$\ \!\! _{\downarrow} U_{\downarrow}$ (with momentum close to 0 or $2
\sqrt{E}$). 
\begin{equation}
\ \!\! _{\downarrow} U_{\mathrm{diag} \downarrow} = 
\sum_{\alpha \sim \beta,\bar{\beta}} \eta_{\alpha} U \eta_{\beta} 
\end{equation}
Then we write 
\begin{eqnarray}
\partial_{t} \bar{G}_{\downarrow} &=& \partial_{t} \bar{G}_{\downarrow}^{(0)} 
  + \partial_{t} \bar{G}_{\downarrow}^{(1)} \\ 
\partial_{t} \bar{G}_{\downarrow}^{(0)} &=& 2t \int \!\! G \ \!\!
  _{\downarrow} \lambda U_{\mathrm{diag} \downarrow} (G-C) \ \!\!
  _{\downarrow} \lambda U_{\downarrow} G \, d\mu(U) \, d\mu(V)
\end{eqnarray}

Keeping only the almost diagonal part for one of the $\lambda U$ allows us to
earn a factor $M^{-r j_{0}/4} = \lambda^{r/2}$ in the small field
region, thanks to lemma \ref{lemprob3}. Therefore, we can treat 
$\partial_{t} \bar{G}_{\downarrow}^{(0)}$ in 
the same way we controlled the higher part and get the following bound. 
\begin{equation}
\| \partial_{t} \bar{G}_{\downarrow}^{(0)} \| \leqslant O(1)
  \lambda^{r/2} \left(\frac{\lambda^{2}}{\sigma}\right)^{2}
  \times \frac{1}{\sigma} 
\end{equation}

We are left with 
\begin{eqnarray}
\partial_{t} \bar{G}_{\downarrow}^{(1)} &=& 2t \lambda^{2} 
  \sum_{{\alpha_{1} \not \sim \alpha_{2}, \bar{\alpha}_{2}} \atop {\alpha_{3},
  \alpha_{4}}} 
  \int \!\! dx \, dy \, \xi(x, y) \nonumber \\ 
  && \qquad \left<G \eta_{\alpha_{1}}(., x) \eta_{\alpha_{2}}(x, .)
    (G-C) \eta_{\alpha_{3}}(., y) \eta_{\alpha_{4}}(y, .) G \right> \\
  &=& 2t \lambda^{2} 
  \sum_{{\alpha_{1} \not \sim \alpha_{2}, \bar{\alpha}_{2}} \atop 
    {\alpha_{3}, \alpha_{4}}} \int \!\! du \, \xi(u) 
    \int \!\! dz \nonumber \\ 
  && \hspace{-1em} \left<G \eta_{\alpha_{1}}(., z) \eta_{\alpha_{2}}(z, .)
    (G-C) \eta_{\alpha_{3}}(., z+u) \eta_{\alpha_{4}}(z+u, .) G \right>
\end{eqnarray}

In the following, we will note 
\begin{equation}
\eta^{(u)}_{\gamma}(x, y) = \eta_{\gamma} (x-y+u) = \eta_{\gamma} (x+u, y) =
  \eta_{\gamma} (x, y-u) 
\end{equation}


\subsection{Sector conservation and unfolding} 
If we look at the vertex in momentum space, we have a factor 
\begin{equation}
\hat{\eta}_{\alpha_{1}}(p_{1}) \hat{\eta}_{\alpha_{2}}(p_{2})
\hat{\eta}_{\alpha_{3}}(p_{3}) \hat{\eta}_{\alpha_{4}}(p_{4})
\delta(p_{1}-p_{2}+p_{3}-p_{4}) 
\end{equation}
This leads to one of the following possibilities (recall that we defined 
$\simeq$ in equation~\ref{defsimeq}):
\begin{equation}
\left\{
  \begin{array}{c}
\alpha_{1} \simeq \alpha_{4} \textrm{ and } \alpha_{2} \simeq \alpha_{3} \\
\textrm{or} \\
\alpha_{1} \simeq \bar{\alpha}_{3} \textrm{ and } \alpha_{2} \simeq
\bar{\alpha}_{3} 
  \end{array}
\right.
\end{equation}
Thus 
\begin{eqnarray}
\partial_{t} \bar{G}_{\downarrow}^{(1)} &=& 2t (I_{0}+J_{0}) \\
I_{0} &=& \lambda^{2} 
  \sum_{\alpha \not \sim \beta, \bar{\beta}} \sum_{{\alpha' \simeq
      \alpha} \atop {\beta' \simeq \beta}}  
  \int \!\! \xi(u) \, du \int \!\! dz 
  \left<G \eta_{\alpha} \eta_{\beta} (G-C) 
    \eta_{\bar{\alpha}'}^{(-u)} \eta_{\bar{\beta}'}^{(u)} G \right> \\
J_{0} &=& \lambda^{2} 
  \sum_{\alpha \not \sim \beta, \bar{\beta}} \sum_{{\alpha' \simeq
      \alpha} \atop {\beta' \simeq \beta}}  
  \int \!\! \xi(u) \, du \int \!\! dz 
  \left<G \eta_{\alpha} \eta_{\beta} (G-C) 
    \eta_{\beta'}^{(-u)} \eta_{\alpha'}^{(u)} G \right> 
\end{eqnarray}

Now we can unfold the $\alpha \beta \bar{\alpha'} \bar{\beta'}$ term, 
\emph{cf.} equations (\ref{eqabab}), (\ref{eqretournement}). 
\begin{equation}
I_{0} = \lambda^{2} 
  \sum_{\alpha \not \sim \beta, \bar{\beta}} \sum_{{\alpha' \simeq
      \alpha} \atop {\beta' \simeq \beta}}  
  \int \!\! \xi(u) \, du \int \!\! dz 
  \left<G \eta_{\alpha} \eta_{\alpha'}^{(u)} (G-C) 
    \eta_{\bar{\beta}} \eta_{\bar{\beta}'}^{(u)} G \right> 
\end{equation}

In order to decouple the $\alpha \alpha'$ and $\beta \beta'$ operators, we
insert 
\begin{equation}
1 = \int \! \delta(z-z') \, dz' = \int \!\! dk \, dz' e^{ik(z-z')}
\end{equation} 
Let us note $e^{ik.}$ the operator whose kernel is 
\begin{equation}
\left(e^{ik.}\right)(x, y) = e^{ikx} \delta(x-y)
\end{equation}
with this notation, we have 
\begin{eqnarray}
I_{0} &=& \lambda^{2} \sum_{\alpha \not \sim \beta, \bar{\beta}}
  \sum_{{\alpha' \simeq \alpha} \atop {\beta'\simeq \beta}} \int \!\! 
  \xi(u) \, du \int \!\! dk \nonumber \\
&& \left<G \eta_{\alpha} e^{ik.} \eta_{\alpha'}^{(u)} (G-C) 
  \eta_{\bar{\beta}} e^{-ik.} \eta_{\bar{\beta}'}^{(u)} G \right>\\ 
J_{0} &=& \lambda^{2} \sum_{\alpha \not \sim \beta, \bar{\beta}}
  \sum_{{\alpha' \simeq \alpha} \atop {\beta'\simeq \beta}} \int \!\! 
  \xi(u) \, du \int \!\! dk \nonumber \\ 
&& \left< \tr \left[ \eta_{\beta} (G-C) \eta_{\beta'}^{(-u)} e^{-ik.} \right] 
  \times G \eta_{\alpha} e^{ik.} \eta_{\alpha'}^{(u)} G \right>  
\end{eqnarray}

We can adopt the following diagrammatic representation 
\begin{eqnarray}
I_{0} &=& 
\begin{picture}(75,25)
\put(0,0){\psfig{figure=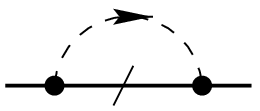}}
\put(9,-7){$\scriptstyle \alpha$} 
\put(20,-7){$\scriptstyle \alpha'$}
\put(49,-7){$\scriptstyle \bar{\beta}$}
\put(61,-7){$\scriptstyle \bar{\beta}'$}
\end{picture}
\\
J_{0} &=& 
\begin{picture}(40,60)
\put(0,0){\psfig{figure=tadpole_in.eps,height=50pt}}
\put(12,-7){$\scriptstyle \alpha$} 
\put(28,-7){$\scriptstyle \alpha'$}
\put(28,12){$\scriptstyle \beta$}
\put(12,12){$\scriptstyle \beta'$}
\end{picture}
\end{eqnarray}

Let us write explicitly the various incoming momenta at the half-vertices 
\begin{eqnarray}
I_{0} &=& \lambda^{2} \sum_{\alpha \not \sim \beta, \bar{\beta}}
  \sum_{{\alpha' \simeq \alpha} \atop {\beta'\simeq \beta}} \int \!\! 
  \xi(u) \, du \int \!\! d\mu(V) \int \!\! dk \int \!\! dw_{1} \ldots dw_{4} 
  \, dp_{1} \ldots dp_{4} \nonumber \\
&& G (., w_{1}) e^{ip_{1} w_{1}} \hat{\eta}_{\alpha}(p_{1}) 
  \hat{\eta}_{\alpha'}(p_{2}) \delta(k-p_{1}+p_{2}) e^{ip_{2}(u-w_{2)}}
  (G -C)(w_{2}, w_{3}) \nonumber \\
&& e^{ip_{3} w_{3}} \hat{\eta}_{\bar{\beta}} (p_{3}) 
  \hat{\eta}_{\bar{\beta}'} (p_{4}) \delta(p_{4}-p_{3}-k) e^{ip_{4}(u-w_{4})}  
  G(w_{4}, .)
  \label{momconserv}
\end{eqnarray}

We can see on equation (\ref{momconserv}) that $k$ allows to go from the
sector $\bar{\beta}$ to the neighboring sector $\bar{\beta}'$. This implies
that $k$ is restricted to a small effective domain around the origin which is
a tube $\zeta_{\beta}$ whose axis is orthogonal to $\beta$ and of size 
$O(1) M^{-j_{1}} \times O(1) M^{-j_{0}/2}$. In the same way, $k$ goes from 
$\alpha$ to $\alpha'$, thus it must be in the intersection of both tubes 
$\zeta_{\alpha}$ and $\zeta_{\beta}$. Since the angle between those tubes is
at least $M^{-r j_{0}/2}$, $k$ has a norm which is at most 
$O(1) M^{r j_{0}/2} M^{-j_{1}}$. 

Thus, we can freely insert a factor $\zeta_{0}(k)$ which restricts the
integration on $k$ to the ball of radius $O(1) M^{r j_{0}/2}
M^{-j_{1}}$. Of course, the same analysis applies to $J_{0}$. 
\begin{eqnarray}
I_{0} &=& \lambda^{2} \sum_{\alpha \not \sim \beta, \bar{\beta}}
  \sum_{{\alpha' \simeq \alpha} \atop {\beta'\simeq \beta}} \int \!\! 
  \xi(u) \, du \int \!\! \zeta_{0}(k) \, dk \nonumber \\
&& \left<G \eta_{\alpha} e^{ik.} \eta_{\alpha'}^{(u)} (G-C) 
  \eta_{\bar{\beta}} e^{-ik.} \eta_{\bar{\beta}'}^{(u)} G \right>\\ 
J_{0} &=& \lambda^{2} \sum_{\alpha \not \sim \beta, \bar{\beta}}
  \sum_{{\alpha' \simeq \alpha} \atop {\beta'\simeq \beta}} \int \!\! 
  \xi(u) \, du \int \!\! \zeta_{0}(k) \, dk \nonumber \\ 
&& \left< \tr \left[ \eta_{\beta} (G-C) \eta_{\beta'}^{(-u)} e^{-ik.}\right] 
  G \eta_{\alpha} e^{ik.} \eta_{\alpha'}^{(u)} G \right>  
\end{eqnarray}

The point is that momentum conservation can be used in a much more efficient
way. If we suppose, in equation (\ref{momconserv}), that the momenta $p_{1}$,
\ldots, $p_{4}$ are in the very low slice $\bar{\Sigma}_{j_{2}}$,
then $k$ will be at the intersection of two tubes of width $M^{-j_{2}}$ and
thus of norm less than $O(1) M^{r j_{0}/2} M^{-j_{2}}$. 

We can introduce 
\begin{equation}
\zeta_{0} = \zeta_{1} + \zeta_{0} (1-\zeta_{1}) = \zeta_{1} +
(\zeta_{0}-\zeta_{1}) 
\end{equation}
where $\zeta_{1}$ forces $|k| \leqslant O(1) M^{r j_{0}/2} M^{-j_{2}}$. This
will give two terms for $I_{0}$ (and $J_{0}$): 
\begin{itemize}
\item a term $I_{1}$ (or $J_{1}$) with $\zeta_{1}$ having a very small 
  transfer momentum $k$ 
\item a term $I_{2}$ (or $J_{2}$) with $\zeta_{0} (1-\zeta_{1})$ which 
  \emph{must} have a leg above the scale $j_{2}$. 
\end{itemize}
\begin{eqnarray}
I_{0} &=& I_{1}+I_{2} \\
I_{1} &=& \lambda^{2} \sum_{\alpha \not \sim \beta, \bar{\beta}}
  \sum_{{\alpha' \simeq \alpha} \atop {\beta'\simeq \beta}} \int \!\! 
  \xi(u) \, du \int \!\! \zeta_{1}(k) \, dk \nonumber \\
&& \left<G \eta_{\alpha} e^{ik.} \eta_{\alpha'}^{(u)} (G-C) 
  \eta_{\bar{\beta}} e^{-ik.} \eta_{\bar{\beta}'}^{(u)} G \right> 
\end{eqnarray}
Let $\eta_{\alpha} = \eta_{\alpha \uparrow \downarrow} +
\eta_{\alpha \downarrow \downarrow}$, 
with $\eta_{\alpha \uparrow \downarrow}$ having its support above the scale
$j_{2}$. 
\begin{eqnarray}
I_{2} &=& \lambda^{2} \sum_{(i_{1}, \ldots, i_{4}) \neq (\downarrow
  \downarrow, \ldots, \downarrow \downarrow)} 
  \sum_{\alpha \not \sim \beta, \bar{\beta}}
  \sum_{{\alpha' \simeq \alpha} \atop {\beta'\simeq \beta}} \int \!\! 
  \xi(u) \, du \int \!\! (\zeta_{0}-\zeta_{1})(k) \, dk \nonumber \\
&& \left<G \eta_{\alpha i_{1}} e^{ik.} \eta_{\alpha' i_{2}}^{(u)} (G-C) 
  \eta_{\bar{\beta} i_{3}} e^{-ik.} \eta_{\bar{\beta}' i_{4}}^{(u)} G\right> 
\end{eqnarray}
In the following, we will forget the indices $i_{1}, \ldots, i_{4}$ for
shortness. 

The terms $I_{2}$ and $J_{2}$ have a high leg, so we treat them by an analogue
of lemma \ref{lemhighpart}. 
\begin{lemma}
\label{lemIiiJii}
\begin{equation}
\|I_{2}\|, \|J_{2}\| \leqslant O(1) (\log \lambda^{-1}) \lambda^{\nu_{2}}
  \left(\frac{\lambda^{2}}{\sigma}\right)^{2} \times \frac{1}{\sigma}
\end{equation}
\end{lemma}
\begin{proof}
Again, we write the covariance as two insertions of an auxiliary field. 
Thanks to sector conservation, we have  
\begin{eqnarray}
I_{2} &=& \sum_{|\theta| \leqslant K M^{-j_{0}/2}} I_{3}(\theta) \\
I_{3}(\theta) &=& \lambda^{2} \sum_{\alpha \not \sim \beta,
  \bar{\beta}} \sum_{\gamma \delta} \int \!\! \xi(u) \, du \int \!\! 
  dw \left[\hat{\zeta}_{0}(w) - \hat{\zeta}_{1}(w) \right] \nonumber \\
&& \left< G \eta_{\alpha} \eta_{\beta }^{(w)} (G-C) 
  \eta_{(\bar{\gamma}+\theta)}^{(-u)} \eta_{\bar{\delta}}^{(w+u)} G 
  \right> \delta_{\alpha \gamma}
\end{eqnarray}
Then we write
\begin{equation}
\delta_{\alpha \gamma} = \left<\omega_{\alpha} \omega_{\gamma} 
  \right>_{d \mu_{\delta}(\boldsymbol{\omega})}, 
\end{equation}
introducing a Gaussian random vector $\boldsymbol{\omega}$. 
\begin{eqnarray}
I_{3}(\theta) &=& \lambda^{2} \int \! \left[\hat{\zeta}_{0}(w) -
  \hat{\zeta}_{1}(w)\right] dw \int \!\! d\mu(V) \, 
  d\mu(U) \, d\mu_{\delta}(\boldsymbol{\omega}) \nonumber \\
&& G \left(\sum_{\alpha \not \sim \beta} \eta_{\alpha}
  \omega_{\alpha} U \eta_{\beta}^{(w)}\right) (G-C) 
  \left( \sum_{\gamma \delta} \eta_{\gamma} 
  \omega_{\bar{\gamma} -\theta} U \eta_{\delta}^{(w)} \right) G 
\end{eqnarray}
Now, we can perform a large field - small field decomposition and stairway
expansions in the small field region. The leading order term corresponds to
the insertion of $\lambda (\boldsymbol{\omega}*U^{j_{0} j_{0}})$ and 
$\lambda ^{2} (\boldsymbol{\omega}*U^{j_{0} j_{2}}) C^{j_{2}}
V^{j_{2} j_{0}}$. This gives a factor 
\begin{equation}
O(1) \sup|\omega_{\alpha}|^{2} \times \lambda^{2} \times 
\lambda^{2} M^{-\frac{(j_{0}-j_{2})}{2}}
\end{equation}
Therefore, we need to control $\left<\sup|\omega_{\alpha}|^{2} \right>_{d
  \mu_{\delta}(\boldsymbol{\omega})}$ in order to conclude. 
This is done with the following lemma. 
\begin{lemma}
Let $\boldsymbol{\omega} \in \RR^{N}$ be a centered Gaussian random vector
  with covariance 
\begin{equation}
\left<\omega_{\alpha} \omega_{\beta}\right>_{d 
  \mu_{\delta}(\boldsymbol{\omega})} = \delta_{\alpha \beta} 
\end{equation}

There exists a constant $C_{0}$ such that for all $N\geqslant 2$ 
\begin{equation}
\left<\sup_{\alpha} |\omega_{\alpha}|^{2}\right>_{d 
\mu_{\delta}(\boldsymbol{\omega})} \leqslant C_{0} \log N
\end{equation}
\end{lemma}

\begin{proof}
Using H{\"o}lder's inequality 
\begin{eqnarray}
\left<\sup_{\alpha} |\omega_{\alpha}|^{2}\right>_{d 
  \mu_{\delta}(\boldsymbol{\omega})} &\leqslant& 
  \left<(\sup_{\alpha} |\omega_{\alpha}|)^{2p}\right>_{d 
  \mu_{\delta}(\boldsymbol{\omega})}^{1/p} \left<1^{q}\right>_{d 
  \mu_{\delta}(\boldsymbol{\omega})}^{1/q} \\ 
&\leqslant& \left<\sum_{\alpha=1}^{N} |\omega_{\alpha}|^{2p}\right>_{d 
  \mu_{\delta}(\boldsymbol{\omega})}^{1/p} \\ 
&\leqslant& N^{1/p} \left[(2p-1)!!\right]^{1/p} \leqslant 2 N^{1/p} (p!)^{1/p}
\end{eqnarray}
Then we take $p=[\log N]$. 
\end{proof}

Thus introducing the vector $\boldsymbol{\omega}$ costs only $\log
\lambda^{-1}$. Then $J_{2}$ can be treated in the same
way, completing the proof.  
\end{proof}

We return now to the bound on $I_{1}$ and $J_{1}$
\begin{lemma}
\label{lemIun}
\begin{equation}
\|I_{1}\| \leqslant O(1) (\log \lambda^{-1}) \lambda^{1-r-4\nu_{2}}
  \left(\frac{\lambda^{2}}{\sigma}\right)^{2} \times \frac{1}{\sigma} 
\end{equation}
\end{lemma}
\begin{proof}
Again we have to get rid of the constraint between $\alpha$ and $\beta$. We
write 
\begin{eqnarray}
\sum_{\alpha \not \sim \beta} f_{\alpha, \beta} &=& \sum_{\alpha \beta}
  f_{\alpha, \beta} - \sum_{\alpha \sim \beta} (f_{\alpha, \beta} + 
  f_{\alpha, \bar{\beta}}) \\
&=& \sum_{\alpha \beta} f_{\alpha, \beta} -\sum_{|\theta| < M^{-r j_{0}/2}}
  \sum_{\alpha \beta} (f_{\alpha, \beta+\theta} + f_{\alpha, 
  \bar{\beta}+\theta}) \delta_{\alpha \beta} \\
&=& \sum_{\alpha \beta} f_{\alpha \beta} -\sum_{|\theta| < M^{-r j_{0}/2}}
  \sum_{\alpha \beta} \left< \omega_{\alpha} \omega_{\beta} f_{\alpha, 
  \beta+\theta} + \omega_{\alpha} \omega_{\beta} f_{\alpha, 
  \bar{\beta}+\theta} \right>_{d \mu_{\delta}(\boldsymbol{\omega})}
\end{eqnarray}
Putting into the expression of $I_{1}$, we get 
\begin{eqnarray}
I_{1} &=& I_{4} + \sum_{|\theta| < M^{-r j_{0}/2}} \left[I_{5}(\theta) +
  \bar{I}_{5}(\theta) \right] \\
I_{4} &=& \lambda^{2} \int \!\! \xi(u) \, du \int \! \zeta_{1}(k) \, dk 
  \int \!\! d\mu(V) \nonumber \\ 
&& G \bigg[\sum_{{\alpha} \atop {\alpha' \simeq \alpha}}
  \eta_{\alpha} \, e^{ik.} \eta_{\alpha'}^{(u)} \bigg] (G-C)  
  \bigg[\sum_{{\beta} \atop {\beta' \simeq \beta}} 
  \eta_{\bar{\beta}} \, e^{-ik.} \eta_{\bar{\beta}'}^{(u)} \bigg] G \\
I_{5}(\theta) &=& \lambda^{2} \int \!\! \xi(u) \, du \int \! \zeta_{1}(k) 
  \, dk  \int \!\! d\mu(V) \, d\mu_{\delta}(\boldsymbol{\omega}) 
  \nonumber \\ 
&& G \bigg[\sum_{{\alpha} \atop {\alpha' \simeq \alpha}} 
  \omega_{\alpha} \eta_{\alpha} \, e^{ik.} \eta_{\alpha'}^{(u)} \bigg] (G-C) 
  \bigg[\sum_{{\beta} \atop {\beta' \simeq \beta}} 
  \omega_{\beta} \eta_{\beta+\theta} \, e^{-ik.} \eta_{\beta'+\theta}^{(u)} 
  \bigg] G 
\end{eqnarray}
$\bar{I}_{5}$ is obtained by changing $\beta$ and $\beta'$ into 
$\bar{\beta}$ and $\bar{\beta}'$ in the expression $I_{5}$. 

We get small factors from the coupling constants $\lambda^{2}$ and the
integration volume for $k$ which is $M^{r j_{0}} M^{-2j_{2}}$. On the other
hand, we must pay for the resolvents and an extra 
$M^{(1-r) j_{0}/2} \sup |\omega_{\alpha}|^{2}$
for $I_{5}=\sum I_{5}(\theta)$. Hence 
\begin{eqnarray}
\| I_{4}\| &\leqslant& O(1) \lambda^{2-2r-4\nu_{2}}
\left(\frac{\lambda^{2}}{\sigma} \right)^{2} \times \frac{1}{\sigma} \\ 
\| I_{5}\| &\leqslant& O(1) (\log \lambda^{-1}) \lambda^{1-r-4\nu_{2}}
\left(\frac{\lambda^{2}}{\sigma} \right)^{2} \times \frac{1}{\sigma} 
\end{eqnarray}
\end{proof}

Now, we are left with $J_{1}$ on which we want to apply our Ward-type
identity. But we need $k.k_{\beta'}$ to be large enough. We define 
\begin{equation}
\zeta_{1}(k) = \theta_{\beta'}(k) + \varepsilon_{\beta'}(k),  
\end{equation}
where $\theta_{\beta'}$ restricts the integration on $k$ to the region 
$k.k_{\beta'} \geqslant \lambda^{2+\nu_{3}}$. This leads to 
\begin{eqnarray}
J_{1} &=& J_{4} + J_{5}\\
J_{4} &=& \lambda^{2} \sum_{\alpha \not \sim \beta} \sum_{{\alpha \simeq
  \alpha'} \atop {\beta' \simeq \beta}} \int \!\! \xi(u) \, du \int \!\!
  \varepsilon_{\beta'}(k) J_{{\alpha \alpha'} \atop {\beta \beta'}}(k) \, dk 
  \\ 
J_{{\alpha \alpha'} \atop {\beta \beta'}}(k)&=& 
  \left< \tr \left[\eta_{\beta} (G-C) \eta_{\beta'}^{(-u)} e^{-ik.} \right] 
    G \eta_{\alpha} e^{ik.} \eta_{\alpha'}^{(u)} G \right>
\end{eqnarray}

\begin{lemma}
\label{lemJiv}
\begin{equation}
\|J_{4}\| \leqslant O(1) \lambda^{\nu_{3}-\varepsilon}
\left(\frac{\lambda^{2}}{\sigma}\right)^{2} \times \frac{1}{\sigma}
\end{equation}
\end{lemma}
\begin{proof}
Let us write $J_{4}$ under the following form 
\begin{eqnarray}
J_{4} &=& \sum_{\beta, \beta' \simeq \beta} \int \!\! \xi(u) \, du \int \!\!
  d\mu(V) \nonumber \\ 
&& G \left[\lambda^{2} \int \!\! dz \, \eta_{\alpha}(., z) T_{\beta
  \beta'}^{(u)}(z)
  \eta_{\alpha'} (z, .) \right] G \\ 
T_{\beta \beta'}^{(u)} (z) &=& \int \!\! \hat{\varepsilon}_{\beta'}(z-z') dz'
  \left(\eta_{\beta}^{(z')}, (G-C) \eta_{\beta'}^{(-z'-u)} \right)
\end{eqnarray}
$\eta_{\beta}^{(z')}$ and $\eta_{\beta'}^{(-z'-u)}$ are now to be considered
as functions and no longer as operators. A stairway expansion on $(G-C)$, in 
the small field region, proves that the leading contribution is obtained by 
restricting $\eta_{\beta}$ and $\eta_{\beta'}$ to the very low slice
$j_{0}$. This yields 
\begin{eqnarray}
\|T_{\beta \beta'}^{(u)}\|_{\infty} &\leqslant& O(1) 
\|\eta_{\beta_{j_{0}}}\|^{2}_{2} \, 
\|G-C\| \int \!\! |\hat{\varepsilon}_{\beta'}(x)| \, dx \\ 
&\leqslant& M^{-j_{0}/2} \left(\frac{\lambda^{2}}{\sigma}\right)
\lambda^{\nu_{3}-\varepsilon} 
\label{eqdeckbeta}
\end{eqnarray}
In order to get (\ref{eqdeckbeta}), we used the fact that our model is
restricted to a single cube so that the integration in the direction
$k_{\beta'}$ is on a domain of size $\lambda^{-2-\varepsilon}$ instead of 
the decaying scale $\lambda^{-2-\nu_{3}}$ of $\hat{\varepsilon}_{\beta'}$. 
The desired bound follows easily. 
\end{proof}

The previous lemma would extend when we study the thermodynamic limit. In that
case, when we work in a given cube $\Delta$, $V$ is replaced by the
corresponding $V_{\Delta}$ whose covariance is 
\begin{equation}
\xi_{\Delta} = \xi^{1/2} \chi_{\Delta} \xi^{1/2} 
\end{equation} 
The set of all $\chi_{\Delta}$ is a partition of unity and each 
$\chi_{\Delta}$ is a smooth function with compact support around the
corresponding cube $\Delta$ \cite{Poi1}. Then we can introduce, 
$\chi_{\bar{\Delta}}$ smooth with compact support around the support
$\bar{\Delta}$ of $\chi_{\Delta}$ and equal to 1 for all points whose distance
to $\bar{\Delta}$ is less than $\lambda^{-2-\varepsilon}$. 

Lemma \ref{lemJiv} can be extended provided we put further localization 
functions at the very beginning of the 
expansion. In the expression of $\partial \bar{G}_{\downarrow}$, equation
(\ref{defGbas}), we can replace the vertex function  
\begin{equation}
\Gamma_{\Delta}(u_{1}, \ldots u_{4}) = \int \!\! \xi_{\Delta}(x, y) \, 
  dx \, dy \, \eta_{\downarrow}(u_{1}, x) \eta_{\downarrow}(x, u_{2}) 
  \eta_{\downarrow}(u_{3}, y) \eta_{\downarrow}(y, u_{4}) 
\end{equation}
by the following one 
\begin{eqnarray}
\tilde{\Gamma}_{\Delta}(u_{1}, \ldots u_{4}) &=& \int \!\! \xi_{\Delta}(x, y)
  \, dx \, dy \, 
  (\chi_{\bar{\Delta}} \eta_{\downarrow})(u_{1}, x) 
  (\eta_{\downarrow} \chi_{\bar{\Delta}})(x, u_{2}) \nonumber \\  
&& \qquad  (\chi_{\bar{\Delta}} \eta_{\downarrow})(u_{3}, y) 
  (\eta_{\downarrow} \chi_{\bar{\Delta}})(y, u_{4})  
\end{eqnarray}
The error term is very small because of the fast decay of $\eta_{\downarrow}$
on a scale $M^{j_{1}} \ll \lambda^{-2-\varepsilon}$ and the 
functions $\chi_{\bar{\Delta}}$ force the tadpole to stay in a cube close to
$\Delta$.


\subsection{Ward term}
Finally, we must deal with 
\begin{equation}
J_{5} = \lambda^{2} \sum_{\alpha \not \sim \beta} \sum_{{\alpha \simeq
  \alpha'} \atop {\beta' \simeq \beta}} \int \!\! \xi(u) \, du \int \!\!
  \theta_{\beta'}(k) J_{{\alpha \alpha'} \atop {\beta \beta'}}(k) \, dk 
\end{equation}

We set 
\begin{equation}
C_{t} = \frac{1}{p^{2}-E-i\sigma-(1-t^{2}) \Sigma}
\end{equation}
Our Ward-type identity relies on the identity 
\begin{eqnarray}
2 k.k_{\beta'} &=& (p+k)^{2} - p^{2}  -2 k.(p-k_{\beta'}) -k^{2} \\
  &=& C_{t}(p+k)^{-1} - C_{t}(p)^{-1} -2 k.(p-k_{\beta'}) -k^{2}
  \nonumber \\
  && \qquad -(1-t^{2}) \left[ \Sigma(p+k)-\Sigma(p)\right]
\label{eqdefward}
\end{eqnarray}

In momentum space, the tadpole insertion of $J_{{\alpha \alpha'} \atop {\beta
    \beta'}}$ can be written 
\begin{equation}
\mathcal{T}(k) = 
\tr \left[\eta_{\beta'}^{(-u)} e^{-ik.} \eta_{\beta} (G-C) \right] = \int \!\!
dp \, \eta_{\beta'}(p) e^{-ipu} \eta_{\beta}(p+k) (G-C) (p+k, p) 
\end{equation}
We insert equation (\ref{eqdefward}) to get  
\begin{eqnarray}
(2 k.k_{\beta'}) \mathcal{T}(k) &=& \tr \left[\eta_{\beta'}^{(-u)} 
  e^{-ik.} \eta_{\beta} C_{t}^{-1} (G-C) \right] - \tr \left[ 
   \eta_{\beta'}^{(-u)} e^{-ik.} \eta_{\beta} 
  (G-C) C_{t}^{-1}\right] \nonumber \\ 
  && - 2k. \tr \left[\eta_{\beta} (G-C) D_{\beta'}^{(-u)} e^{-ik.} \right] 
   - k^{2} \mathcal{T}(k) \nonumber \\
  && +(1-t^{2}) \, \tr \left[\eta_{\beta} (G-C) \eta_{\beta'}^{(-u)} 
    (\Sigma e^{-ik.} - e^{-ik.} \Sigma) \right] \\
D_{\beta'}(x, y) &=& \int \! dp \, e^{ip(x-y)} (p-k_{\beta'}) \eta_{\beta'}(p) 
\end{eqnarray}

Now, we use the resolvent identity 
\begin{equation}
G = C_{t} - C_{t} \lambda V G = C_{t} - G \lambda V C_{t}
\end{equation}
Since $C$ and $C_{t}$ commute, we get 
\begin{eqnarray}
(2 k.k_{\beta'}) \mathcal{T}(k) &=& - \tr \left[\eta_{\beta'}^{(-u)}
  e^{-ik.} \eta_{\beta} \lambda V G \right] + \tr \left[G \lambda V 
  \eta_{\beta'}^{(-u)} e^{-ik.} \eta_{\beta} \right] \nonumber \\ 
  && - 2k. \tr \left[\eta_{\beta} (G-C) D_{\beta'}^{(-u)} e^{-ik.} \right] 
  -k^{2} \mathcal{T}(k) \nonumber \\
  && +(1-t^{2}) \, \tr \left[\eta_{\beta} (G-C) \eta_{\beta'}^{(-u)} 
    (\Sigma e^{-ik.} - e^{-ik.} \Sigma) \right] 
  \label{eqtermesward}
\end{eqnarray}

We put (\ref{eqtermesward}) back into the expression of $J_{5}$, writing 
\begin{equation}
J_{5} = -J_{R} + J_{L} - J_{D} - J_{k^{2}} +J_{\Sigma}
\end{equation}
where the notations refer directly to the various terms of equation 
(\ref{eqtermesward}). 

\begin{lemma}
\label{lempetitsJ}
\begin{eqnarray}
\| J_{D} \| &\leqslant& O(1) (\log \lambda^{-1})
\lambda^{1-r-2\nu_{2}-\varepsilon} 
\left(\frac{\lambda^{2}}{\sigma}\right)^{2} \times \frac{1}{\sigma} \\ 
\| J_{k^{2}} \| &\leqslant& O(1) (\log \lambda^{-1})
\lambda^{2-2r-4\nu_{2}-\varepsilon}  
\left(\frac{\lambda^{2}}{\sigma}\right)^{2} \times \frac{1}{\sigma} \\ 
\| J_{\Sigma} \| &\leqslant& O(1) (\log \lambda^{-1})
\lambda^{2-r-2\nu_{2}-\varepsilon}  
\left(\frac{\lambda^{2}}{\sigma}\right)^{2} \times \frac{1}{\sigma}
\end{eqnarray}
\end{lemma}
\begin{proof}
We can treat $J_{D}$, $J_{k^{2}}$ and $J_{\Sigma}$ the way we treated 
$J_{4}$ because we have earned small factors: 
\begin{itemize}
\item for $J_{D}$, we earn something thanks to 
\begin{equation}
k.D_{\beta'} \sim |k| M^{-j_{0}/2} \eta_{\beta'} \sim M^{-(1-r) j_{0}/2}
M^{-j_{2}} 
\end{equation}
but we have still the spatial integration of the tadpole to pay, which costs 
\begin{equation}
\int \!\! dx \, \left| \widehat{
\left(\frac{\theta_{\beta'}}{2k.k_{\beta'}}\right)} (x) \right| \leqslant
O(1) (\log \lambda^{-1}) \lambda^{-2-\varepsilon} 
\end{equation}
This is because we have 
\begin{eqnarray}
\left\|\widehat{\frac{\theta_{\beta'}}{2k.k_{\beta'}}} \right\|_{\infty}
  &\leqslant& O(1) M^{r \frac{j_{0}}{2}} M^{-j_{2}} 
  \int_{\lambda^{2+\nu_{3}}}^{M^{r \frac{j_{0}}{2}} M^{-j_{2}}}
  \frac{dk_{/\!\!/}}{2 |k_{\beta'}| k_{/\!\!/}} \\ 
&\leqslant& O(1) M^{r \frac{j_{0}}{2}} M^{-j_{2}} \log \lambda^{-1}, 
\end{eqnarray}
and the spatial integration is in a volume $O(1) M^{-r \frac{j_{0}}{2}} 
M^{+j_{2}} \times \lambda^{-2-\varepsilon}$. 

Therefore  
\begin{equation}
\| J_{D} \| \leqslant O(1) (\log \lambda^{-1}) \lambda^{1-r-2\nu_{2}}
\left(\frac{\lambda^{2}}{\sigma}\right)^{2} \times \frac{1}{\sigma}
\end{equation} 
\item for $J_{k^{2}}$, we earn something from 
\begin{equation}
|k^{2}| \leqslant M^{r j_{0}} M^{-2j_{2}} 
\end{equation}
and the spatial integration of the tadpole has the same price as before. 
\begin{equation}
\| J_{D} \| \leqslant O(1) (\log \lambda^{-1})
\lambda^{2-2r-4\nu_{2}-\varepsilon} 
\left(\frac{\lambda^{2}}{\sigma}\right)^{2} \times \frac{1}{\sigma}
\end{equation}
\item finally, for $J_{\Sigma}$, we notice that $\Sigma$ is an almost local
  operator whose norm is proportional to $\lambda^{2}$. Thus, taking the 
  commutator with $e^{-ik.}$ gives a gradient term which is very small 
\begin{equation}
\|[\Sigma, e^{-ik.}]\| \leqslant O(1) \lambda^{2} |k| 
\end{equation}
We can conclude
\begin{equation}
\| J_{\Sigma} \| \leqslant O(1) (\log \lambda^{-1})
\lambda^{2-r-2\nu_{2}-\varepsilon} 
\left(\frac{\lambda^{2}}{\sigma}\right)^{2} \times \frac{1}{\sigma}
\end{equation} 
\end{itemize}
\end{proof}

We are left with $-J_{R}+J_{L}$ 
\begin{eqnarray}
-J_{R} &=& -\lambda^{2} \sum_{\alpha \not \sim \beta, \bar{\beta}}
\sum_{{\alpha' \simeq \alpha} \atop {\beta' \simeq \beta}} \int \!\! \xi(u) \,
du \int \!\! \frac{\theta_{\beta'}(k)}{2k.k_{\beta'}} \, dk \nonumber \\ 
&& \left< \tr\left[\eta_{\beta'}^{(-u)} e^{-ik.} \eta_{\beta} 
    \lambda V G \right] G \eta_{\alpha} e^{ik.} \eta_{\alpha'}^{(u)} 
  \right> \\ 
J_{L} &=& \lambda^{2} \sum_{\alpha \not \sim \beta, \bar{\beta}}
\sum_{{\alpha' \simeq \alpha} \atop {\beta' \simeq \beta}} \int \!\! \xi(u) \,
du \int \!\! \frac{\theta_{\beta'}(k)}{2k.k_{\beta'}} \, dk \nonumber \\ 
&& \left< \tr\left[G \lambda V \eta_{\beta'}^{(-u)} e^{-ik.} 
    \eta_{\beta} \right] G \eta_{\alpha} e^{ik.} \eta_{\alpha'}^{(u)} 
  \right> 
\end{eqnarray}
At this point, we use a diagrammatic representation to perform
the connection with the heuristic presentation of section \ref{secheuristic}. 
It is easy to see that we have 

\vspace{1em}
\begin{equation}
J_{R} = \left<
\begin{picture}(50,20)
\put(0,0){\psfig{figure=graph_C.eps,height=50pt}} 
\put(17,-7){$\scriptstyle \alpha$}
\put(29,-7){$\scriptstyle \alpha'$}
\put(12,16){$\scriptstyle \beta'$}
\put(28,16){$\scriptstyle \beta$}
\end{picture}
\right>
\quad \textrm{and} \quad 
J_{L} = \left<
\begin{picture}(50,20)
\put(0,0){\psfig{figure=graph_D.eps,height=50pt}} 
\put(17,-7){$\scriptstyle \alpha$}
\put(29,-7){$\scriptstyle \alpha'$}
\put(15,16){$\scriptstyle \beta'$}
\put(30,16){$\scriptstyle \beta$}
\end{picture}
\right> 
\end{equation}

We take the degenerate part of the $V$ away 
\begin{equation}
\eta_{\beta} V = \sum _{\gamma \sim \beta, \bar{\beta}} \eta_{\beta} V
\eta_{\gamma} + \sum _{\gamma \not \sim \beta, \bar{\beta}} \eta_{\beta} V
\eta_{\gamma}
\end{equation}
so that we can write 
\begin{equation}
J_{R, L} = J_{R, L}^{(0)} + J_{R, L}^{(1)}. 
\end{equation}
$J_{R, L}^{(0)}$ is the almost diagonal $V$ part, it has a bound 
\begin{equation}
\| J_{R, L}^{(0)} \| \leqslant O(1) (\log \lambda^{-1})
\lambda^{\frac{r}{2}-\varepsilon} 
\left(\frac{\lambda^{2}}{\sigma}\right)^{2} \times \frac{1}{\sigma}
\end{equation}

We integrate the $V$ by parts in $J_{R, L}^{(1)}$, and we use sector
conservation and unfolding to generate the 12 terms of equation 
(\ref{termesward}). 

\begin{lemma}
There exists $\nu>0$ such that 
\begin{equation}
  \| J_{R, L}^{(1)} \| \leqslant O(1) (\log \lambda^{-1})^{3} \lambda^{\nu} 
\left(\frac{\lambda^{2}}{\sigma}\right)^{3} \times \frac{1}{\sigma}
\end{equation}
where 
\begin{eqnarray}
\nu &=& \min \{ (2-r-4\nu_{1}-2\nu_{2}), (\nu_{2}+\frac{r}{2}-\varepsilon),
\nonumber \\ 
&& \quad (1-r-2\nu_{2}-\nu_{3}-\varepsilon), 2(\nu_{1}-\nu_{2}-r)-\varepsilon
\} 
\end{eqnarray}
\end{lemma}

\begin{proof}
Let us bound the various terms of (\ref{termesward}). 
First, we consider the graphs without loops. We explain the bound for a
typical one 

\vspace{1em}
\begin{equation}
\mathcal{A}_{1} = 
\left<
\begin{picture}(52,20)
\put(0,-1){\psfig{figure=graph_C56.eps,width=50pt}}
\put(26,31){$\scriptstyle \beta$} 
\put(40,-7){$\scriptstyle \beta''$}
\end{picture}
\right>
= \left<
\begin{picture}(76,20)
\put(0,-17){\psfig{figure=graph_C5.eps,width=75pt}} 
\put(54,8){$\scriptstyle \beta$}
\end{picture}
\right>
\end{equation}

One can check that the analytic expression for $\mathcal{A}_{1}$ is 
\begin{eqnarray}
\mathcal{A}_{1} &=& \lambda^{4} \sum_{{\alpha \not \sim \beta, \bar{\beta}}
  \atop {\gamma \not \sim \beta, \bar{\beta}}}
  \sum_{{\alpha' \simeq \alpha} \atop {\beta' \simeq \beta}} 
  \sum_{{\beta'' \simeq \beta} \atop {\gamma' \simeq \gamma}} 
  \int \!\! \xi(u) \xi(v) \, du \, dv \int \!\! \frac{\theta_{\beta'}(k)}{2
  k.k_{\beta'}} \zeta_{0}(k') \, dk \, dk' \nonumber \\ 
  && \left< G \eta_{\alpha} e^{ik.} \eta_{\alpha'}^{(u)} G
  \eta_{\gamma'}^{(-v)} e^{ik'.} \eta_{\gamma} G \eta_{\beta'}^{(-u)} e^{-ik.}
  \eta_{\beta} e^{-ik'} \eta_{\beta''}^{(v)} G \right>
\end{eqnarray}
We bound $\mathcal{A}_{1}$ the way we bounded $I_{1}$ in lemma 
\ref{lemIun}. 
\begin{itemize}
\item Our small factors are $\lambda^{4}$ and the integration on 
  $k$ and $k'$, \emph{i.e.} $(\log \lambda^{-1}) M^{r j_{0}/2} M^{-j_{2}}$ 
  and $M^{r j_{0}} M^{-2j_{1}}$. 
\item We must get rid of the constraints on $\alpha$, $\beta$ and
  $\gamma$. This is done by introducing Gaussian random vectors and 
  costs $(M^{(1-r) j_{0}/2} \log \lambda^{-1})^{2}$. Finally we
  must pay for the resolvents. 
\end{itemize}
Gathering all factors, we get 
\begin{equation}
\|\mathcal{A}_{1} \| \leqslant O(1) (\log \lambda^{-1})^{3} \lambda^{2-r-4
  \nu_{1}-2\nu_{2}} \left(\frac{\lambda^{2}}{\sigma}\right)^{3} \times
  \frac{1}{\sigma}  
\end{equation}

Now, let us see how we can pair the graphs with loops to get a small
result. For instance let us consider 
\begin{equation}
\mathcal{A}_{2} = 
\left<
\begin{picture}(35,60)
\put(0,0){\psfig{figure=graph_C3.eps,height=60pt}} 
\put(21,11){$\scriptstyle \beta$} 
\put(7,11){$\scriptstyle \beta'$} 
\put(27,21){$\scriptstyle \beta''$}
\end{picture}
- 
\begin{picture}(35,60)
\put(0,0){\psfig{figure=graph_D3.eps,height=60pt}} 
\put(7,11){$\scriptstyle \beta$}
\put(21,11){$\scriptstyle \beta''$}
\put(0,21){$\scriptstyle \beta'$}
\end{picture} 
\right>
\end{equation}

We know that the momentum $k$ at the first vertex is bounded by 
$K_{1} M^{r j_{0}/2} M^{-j_{2}}$. Therefore, we can find $K_{2}$ such that if
the momentum $k'$ at the second vertex is larger than $K_{2} M^{r j_{0}}
M^{-j_{2}}$ in norm, we have a leg higher than $2 K_{1} M^{r j_{0}/2}$. This
leads to the decomposition 
\begin{equation}
\mathcal{A}_{2} = \mathcal{A}_{2}^{\mathrm{high}} +
\mathcal{A}_{2}^{\mathrm{low}} 
\end{equation}

In $\mathcal{A}_{2}^{\mathrm{high}}$, we have a high leg at the second
vertex. But this leg can be $\eta_{\beta}$ (the thick line) and this would
prevent us from making a staiway expansion and getting a small factor. 
Yet, in that case, we would know that
at the first vertex, $\eta_{\beta'}$ (resp. $\eta_{\beta''}$) had to be higher
than $K_{1} M^{r j_{0}/2} M^{-j_{2}}$. Therefore, in the same way we did in 
lemma \ref{lempetitsJ} we can show 
\begin{equation}
\| \mathcal{A}_{2}^{\mathrm{high}} \| \leqslant O(1) (\log \lambda^{-1}) 
  \lambda^{\nu_{2}+ r/2-\varepsilon} 
  \left(\frac{\lambda^{2}}{\sigma}\right)^{3} \times
  \frac{1}{\sigma}  
\end{equation}

For $\mathcal{A}_{2}^{\mathrm{low}}$, we use the fact that the first graph is
equal to the second up to error terms. 
\begin{itemize}
\item We change $\frac{\theta_{\beta'}(k)}{2 k.k_{\beta'}}$ into  
$\frac{\theta_{\beta}(k)}{2 k.k_{\beta}}$. The remainder term bears a 
factor  
\begin{equation}
|k.(k_{\beta}-k_{\beta'})| \lambda^{-2-\nu_{3}} \leqslant O(1)
\lambda^{1-r-2\nu_{2}-\nu_{3}}  
\end{equation} 
\item We exchange the ends of the two dashed lines in the middle loop. This 
  amounts to commute first $e^{-ik.}$ and $\eta_{\beta}$ and then 
  $e^{-ik'.}$ and $\eta_{\beta}$. Thus the error term has an extra factor 
\begin{equation}
\max (|k|, |k'|) M^{j_{1}} \leqslant O(1) M^{r j_{0}} M^{-j_{2}} M^{j_{1}}
\leqslant O(1) \lambda^{2(\nu_{1}-\nu_{2}-r)}  
\end{equation}
\end{itemize}

In conclusion, we obtain 
\begin{equation}
\| \mathcal{A}_{2}^{\mathrm{low}} \| \leqslant O(1) (\log \lambda^{-1}) 
  \max [\lambda^{1-r-2\nu_{2}-\nu_{3}-\varepsilon}, 
  \lambda^{2(\nu_{1}-\nu_{2}-r)-\varepsilon}] 
  \left(\frac{\lambda^{2}}{\sigma}\right)^{3} \times
  \frac{1}{\sigma}  
\end{equation}
\end{proof}

Taking $\nu_{1}$ small (but not too small) and 
$r$, $\nu_{2}$ and $\nu_{3}$ very small, the various powers of $\lambda$
(standing for the small factors we earned) that we met all along the
demonstration are indeed positive. This concludes the proof of theorem
\ref{thward}.



\begin{thebibliography}{99}
\bibitem{PF} L. Pastur, and A. Figotin, \emph{Spectra of Random and
    Almost-Periodic Operators}, Springer-Verlag (1992). 

\bibitem{Weg} F.J. Wegner, \emph{Phys. Rev.} \textbf{B19} (1979) 783 

\bibitem{OW} R. Opperman, and F.J. Wegner, \emph{Z. Phys.} \textbf{B34}
  (1979), 327. 

\bibitem{CH} J.-M. Combes, and P.D. Hislop, ``Localization for Some
  Continuous Random Hamiltonians in $d$-Dimensions'', \emph{Journ. Func. Anal.}
  \textbf{124} (1994), 149.

\bibitem{Poi1} G. Poirot ``Mean Green's function of the Anderson model 
  at weak disorder with an infra-red cut-off'', (cond-mat/9702111) to appear. 

\bibitem{Poi2} G. Poirot, \emph{Mod{\`e}le d'Anderson {\`a} faible d{\'e}sordre},  
  th{\`e}se de l'{\'E}cole Polytechnique (1998). 

\bibitem{MPR} J. Magnen, G. Poirot, and V. Rivasseau, ``The Anderson Model
  as a Matrix Model'' \emph{Nucl. Phys. B (Proc. Suppl.)}, \textbf{58} 
  (1997), 149. 

\bibitem{FMRT1} J. Feldman, J. Magnen, V. Rivasseau, and E. Trubowitz, 
  in \emph{The State of Matter}, World Scientific (1994), 293.

\bibitem{FMRT2} J.Feldman, J. Magnen, V. Rivasseau, and E. Trubowitz,
  \emph{Europhys. Lett.} {\bf 24} (1993), 521.

\end{thebibliography}
\end{document}